\documentclass[twocolumn]{aastex63}
\hypersetup{linkcolor=blue,urlcolor=blue,citecolor=blue,filecolor=cyan}

\shorttitle{A Massive Protocluster at $z=6.6$}
\shortauthors{Wang et al.}
\graphicspath{{./}{figures/}}

\begin{document}

\title{A Massive Protocluster Anchored by a Luminous Quasar at $z=6.63$}

\correspondingauthor{Feige Wang}
\email{feigewang@arizona.edu}

\author[0000-0002-7633-431X]{Feige Wang}
\affiliation{Steward Observatory, University of Arizona, 933 N Cherry Avenue, Tucson, AZ 85721, USA}

\author[0000-0001-5287-4242]{Jinyi Yang}
\affiliation{Steward Observatory, University of Arizona, 933 N Cherry Avenue, Tucson, AZ 85721, USA}

\author[0000-0002-7054-4332]{Joseph F.\ Hennawi}
\affiliation{Department of Physics, University of California, Santa Barbara, CA 93106-9530, USA}
\affiliation{Leiden Observatory, Leiden University, Niels Bohrweg 2, NL-2333 CA Leiden, Netherlands}

\author[0000-0003-3310-0131]{Xiaohui Fan}
\affiliation{Steward Observatory, University of Arizona, 933 N Cherry Avenue, Tucson, AZ 85721, USA}

\author[0000-0002-5367-8021]{Minghao Yue}
\affil{Steward Observatory, University of Arizona, 933 North Cherry Avenue, Tucson, AZ 85721, USA}

\author[0000-0002-2931-7824]{Eduardo Ba\~nados}
\affil{Max Planck Institut f\"ur Astronomie, K\"onigstuhl 17, D-69117, Heidelberg, Germany}

\author{Shane Bechtel}
\affiliation{Department of Physics, University of California, Santa Barbara, CA 93106-9530, USA}

\author[0000-0002-1620-0897]{Fuyan Bian}
\affil{European Southern Observatory, Alonso de C\'ordova 3107, Casilla 19001, Vitacura, Santiago 19, Chile}

\author[0000-0001-8582-7012]{Sarah Bosman}
\affiliation{Max Planck Institut f\"ur Astronomie, K\"onigstuhl 17, D-69117, Heidelberg, Germany}

\author[0000-0002-6184-9097]{Jaclyn B. Champagne}
\affiliation{Steward Observatory, University of Arizona, 933 N Cherry Avenue, Tucson, AZ 85721, USA}

\author[0000-0003-0821-3644]{Frederick B. Davies}
\affil{Max Planck Institut f\"ur Astronomie, K\"onigstuhl 17, D-69117, Heidelberg, Germany}

\author[0000-0002-2662-8803]{Roberto Decarli}
\affil{INAF--Osservatorio di Astrofisica e Scienza dello Spazio, via Gobetti 93/3, I-40129, Bologna, Italy}

\author[0000-0002-6822-2254]{Emanuele Paolo Farina}
\affil{Gemini Observatory, NSF’s NOIRLab, 670 N A’ohoku Place, HI-96720, Hilo, USA}

\author[0000-0002-5941-5214]{Chiara Mazzucchelli}
\affil{Instituto de Estudios Astrof\'{\i}sicos, Facultad de Ingenier\'{\i}a y Ciencias, Universidad Diego Portales, Avenida Ejercito Libertador 441, Santiago, Chile}

\author[0000-0001-9024-8322]{Bram Venemans}
\affiliation{Leiden Observatory, Leiden University, Niels Bohrweg 2, NL-2333 CA Leiden, Netherlands}

\author[0000-0003-4793-7880]{Fabian Walter}
\affil{Max Planck Institut f\"ur Astronomie, K\"onigstuhl 17, D-69117, Heidelberg, Germany}

\begin{abstract}
Protoclusters, the progenitors of galaxy clusters, trace large scale structures in the early Universe and are important to our understanding of structure formation and galaxy evolution. To date, only a handful of protoclusters have been identified in the Epoch of Reionization (EoR). As one of the rarest populations in the early Universe, distant quasars that host active supermassive black holes are thought to reside in the most massive dark matter halos at that cosmic epoch, and could thus potentially pinpoint some of the earliest protoclusters. In this letter, we report the discovery of a massive protocluster around a luminous quasar at $z=6.63$. This protocluster is anchored by the quasar, and includes three 
[\ion{C}{2}] emitters at $z\sim6.63$, 12 spectroscopically confirmed Ly$\alpha$ emitters (LAEs) at $6.54<z\le6.64$, and a large number of narrow-band imaging selected LAE candidates at the same redshift. This structure has an overall overdensity of $\delta=3.3^{+1.1}_{-0.9}$ within $\sim35\times74$ cMpc$^2$ on the sky
and an extreme overdensity of $\delta>30$ in its central region (i.e., $R\lesssim2$ cMpc). We estimate that this protocluster will collapse into a galaxy cluster with a mass of $6.9^{+1.2}_{-1.4}\times10^{15}~M_\odot$ at the current epoch, more massive than the most massive clusters known in the local Universe such as Coma. In the quasar vicinity, we discover a double-peaked LAE which implies that the quasar has a UV lifetime greater than 0.8 Myrs and has already ionized its surrounding intergalactic medium.

\end{abstract}

\keywords{Quasar (1319) --- Protoclusters(1297) --- Supermassive black holes (1663) --- Reionization (1383) --- Early universe (435)}

\section{Introduction} \label{sec:intro}
How luminous quasars, powered by billion-solar-mass supermassive black holes (SMBHs), formed less than a billion years after the Big Bang has been one of the outstanding open questions in cosmology since the early 2000s. In the current theoretical paradigm, the earliest SMBHs grew in massive galaxies and inhabit massive dark matter halos \citep[e.g.,][]{Springel05, Costa14, Latif22}. This picture is supported by the observed strong clustering of high redshift quasars \citep[e.g.,][]{Shen07, Arita23} and the discoveries of a significant number of binary quasars at $z>4$ \citep[e.g.,][]{hennawi06b, McGreer16, Yue21}. Following these arguments, one might expect that the most distant quasars, hosting the earliest active SMBHs, could act as signposts for protoclusters, the progenitors of local galaxy clusters and the largest (well beyond 1 cMpc) structures in the early Universe. 

Extensive efforts on finding overdensities of galaxies around $z>6$ quasars \citep[e.g.,][]{Stiavelli05, Willott05, Zheng06, Kim09, Banados13, Morselli14, Champagne23} have been made in the past two decades \citep[see][for a review]{Overzier16}, however, to date only four $z>6$ quasar fields were spectroscopically confirmed to be associated with galaxy overdensities at Mpc-scales: 
J0100+2802 at $z=6.3$ with $\sim$20 [\ion{O}{3}] emitters confirmed with recent JWST observations \citep{Kashino22},
J0305--3150 at $z=6.6$ with 10 [\ion{O}{3}] emitters and a large number of Lyman Break Galaxies (LBGs) confirmed with recent JWST observations \citep{Wang23a} and Hubble Space Telescope (HST) observations \citep{Champagne23}, respectively,
J1030+0524 at $z=6.3$ with two LAEs and four LBGs at the quasar redshift \citep{Mignoli20} 
and J1526--2050 at $z=6.6$ with two LAEs and two [\ion{C}{2}] emitters \citep{Decarli17, Neeleman19b, Meyer22}. 

 \begin{figure*}[tbh]
\centering
\includegraphics[trim=250 200 420 80, clip, width=1.0\linewidth]{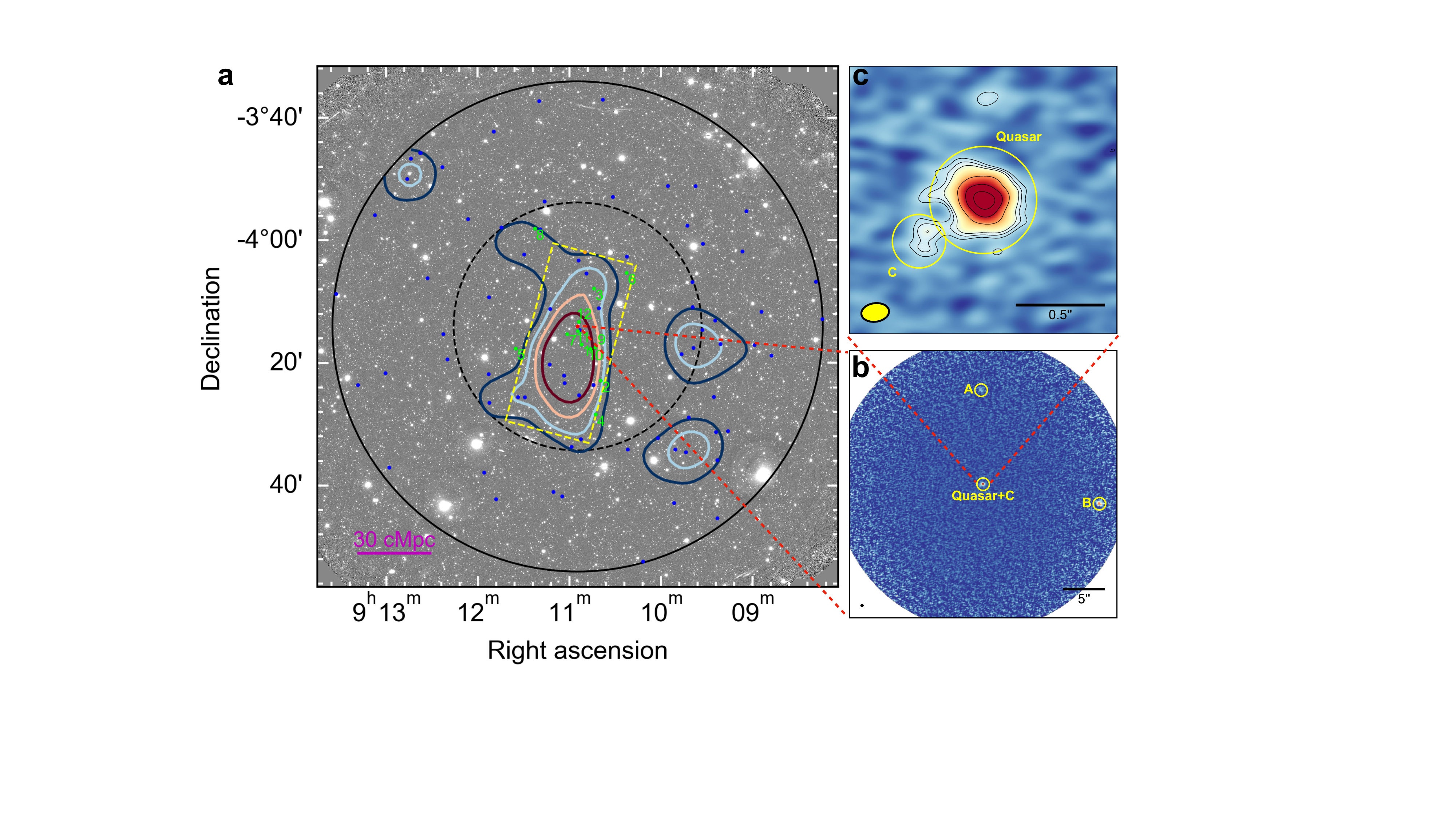}
\caption{\small 
{\bf a,} Large scale overdensity of LAEs around quasar J0910--0414. 
The background image is HSC $z$-band image. 
The red dot denotes the position of the quasar. 
The blue dots represent LAE candidates identified from deep Subaru imaging, 
the green dots represent spectroscopically confirmed LAEs at $z>6.5$, 
and the orange point represents a spectroscopically confirmed low-$z$ galaxy. 
The overdensity of LAEs within the yellow box (14$\times$30 arcmin$^2$, or $\sim35\times74$ cMpc$^2$) is $\delta=4.3^{+1.1}_{-0.9}$.
The colored contours are overdensity isolines of LAEs which were estimated using a quartic kernel and a grid size of 1 arcmin. The blue, light blue, pink, and red lines denote 
overdensity $\delta=3, 4, 5,$ and $6$, respectively. 
The black dashed and solid lines represent $r=50$ cMpc (at $z=6.63$) and $r=40'$ ($\sim 99$ cMpc at $z=6.63$), respectively. The average LAE surface density was estimated using galaxies located between these two lines.
{\bf b,} Small scale overdensity of [\ion{C}{2}] emitting galaxies identified from ALMA observations.
{\bf c,} 
The zoom-in [\ion{C}{2}] flux integrated map of the quasar host galaxy and a satellite galaxy (C) with [\ion{C}{2}] emission. 
}
\label{fig:map}
\end{figure*}

In this letter, we report the discovery of a protocluster anchored by a luminous quasar: J0910--0414 at $z=6.63$. 
The identification of this protocluster is secured by the discovery of a large number of narrow-band imaging selected LAE candidates, twelve spectroscopically confirmed LAEs, and three [\ion{C}{2}] emitters in addition to the central luminous quasar. 
In \S \ref{sec:obs}, we present our observations, data reduction, LAE selection and confirmation, as well as ALMA confirmation of [\ion{C}{2}] emitters. In \S \ref{sec:cluster}, we show the characterization of the overdensity of both LAEs and [\ion{C}{2}] emitters and discuss the constraints on the mass of the protocluster. We also compare the structure anchored by J0910--0414 with other structures traced by distant quasars in this section. In addition, we report the discovery of a double-peaked LAE in the quasar vicinity and its implications in this section. 
Finally, we summarize our findings and discuss future implications in \S \ref{sec:conclusion}.

\section{Observations and data analysis} \label{sec:obs}
J0910--0414, a luminous quasar discovered by \citep{Wang19b} with a [\ion{C}{2}]-based redshift of $z=6.6363\pm0.0003$ \citep{Wang23b}. J0910--0414 hosts a SMBH with a virial mass of $(3.59\pm0.61)\times10^9~M_\odot$ \citep{Yang21}. The black hole in J0910--0414 is among the most massive SMBHs known at $z>6.5$ and therefore is an ideal candidate for tracing the most biased dark matter halos and galaxy overdensities. 
In addition, the redshifted Ly$\alpha$ line of galaxies at the quasar redshift is well placed in a narrow band filter, NB926, on Subaru/HSC, which makes J0910--0414 a good target to search for an overdensity of LAEs with narrow-band imaging. This motivated us to investigate whether J0910--0414 traces an overdensity of galaxies with multi-wavelength followup observations. 

\subsection{Subaru/HSC observation and LAE selection} \label{subsec:hsc}
The Subaru/HSC observations were obtained from 2019 November to 2020 January. We observed this quasar field with $i2$, $z$ and $NB926$ filters 
and the on-source exposure times are 150 minutes, 240 minutes, and 313 minutes, respectively. The data were processed with {\tt hscPipe 6.7}, a customization of the 
LSST Science Pipelines \citep{Juric17,Bosch18}. The pipeline first processes and calibrates the individual CCDs and then combines them into stacked mosaics 
for each filter. The photometric and astrometric calibrations were performed based on the Pan-STARRS1 DR1 dataset \citep{Chambers16}. 
The seeing of stacked images were estimated to be $0.79''$, $0.74''$, and $0.56''$ for $i2$, $z$ and $NB926$, respectively. The 5$\sigma$ limiting magnitudes measured within
$1.5''$ diameter apertures were measured to be 27.31, 26.27, and 25.71 AB magnitudes for $i2$, $z$ and $NB926$, respectively. 
The {\tt hscPipe} pipeline performs a range of photometric measurements. In this work, we used the {\tt forced} catalog following previous works \citep{Becker18,Shibuya18}.
The {\tt forced} catalog contains the intrinsic astrometric parameters of a source that are determined from the band in which it is detected with the highest significance. 
These parameters are then held fixed for other bands. For the photometric information, it contains CModel magnitude, aperture magnitude and a number of pipeline flags. 
The CModel magnitude is a weighted combination of exponential and de Vaucouleurs fits to the light profiles of each object. The detailed algorithm of the CModel 
photometry is presented in \cite{Bosch18}. Except where noted below, we adopt CModel magnitudes for the color and total magnitude measurements.

Subaru/HSC covers a circular field-of-view (FoV) with a radius of $45'$ and our designed dither pattern gives a relative uniform depth within $r\lesssim40'$ from the central quasar position. 
Therefore, the Ly$\alpha$ emitter (LAE) within $40'$ of the quasar position will have a uniform selection. 
Similar with \cite{Shibuya18}, we impose the following selection criteria to select LAEs:
\begin{equation}
SNR(NB926)\ge5
\end{equation}
\begin{equation}
z-NB926\ge1.0
\end{equation}
\begin{equation}
SNR(i)<3
\end{equation}
The signal-to-noise ratio (SNR) is measured in a $1.5''$ diameter aperture while the color was measured using the CModel magnitude following previous works \citep[e.g.,][]{Shibuya18}.
We also required {\tt [i2,z,NB926]\_flag\_edge}, {\tt [i2,z,NB926]\_flag\_bad},  {\tt [i2,z,NB926]\_flag\_intercenter},  
\\
{\tt [i2,z,NB926]\_flag\_saturcenter}, and  
\\
{\tt [i2,z,NB926]\_flag\_crcenter}
to be {\tt False} to reject most of the artifacts and/or low-$z$ contaminants introduced by inaccurate photometry. 
As a usual step on selecting LAEs, all sources satisfying these criteria were then visually inspected in the stacked images to further reject moving objects, satellite trails, uncorrected hot pixels, and other artifacts. 
In total, 87 LAE candidates pass our selection. The spatial distribution of these LAEs is shown in Figure \ref{fig:map}. For visualization purpose, we estimate the overdensity contours of LAEs using a quartic kernel and a $1'$ grid size which are also shown in Figure \ref{fig:map}. Many of these LAEs are distributed within the central $\sim 14\times30$ arcmin$^2$, or $\sim35\times74$ cMpc$^2$ and 
clustered around the central luminous quasar, indicating that the quasar resides in a significant overdense structure. 

 \begin{figure*}[tbh]
\centering
\includegraphics[trim=395 270 150 130, clip, width=1\linewidth]{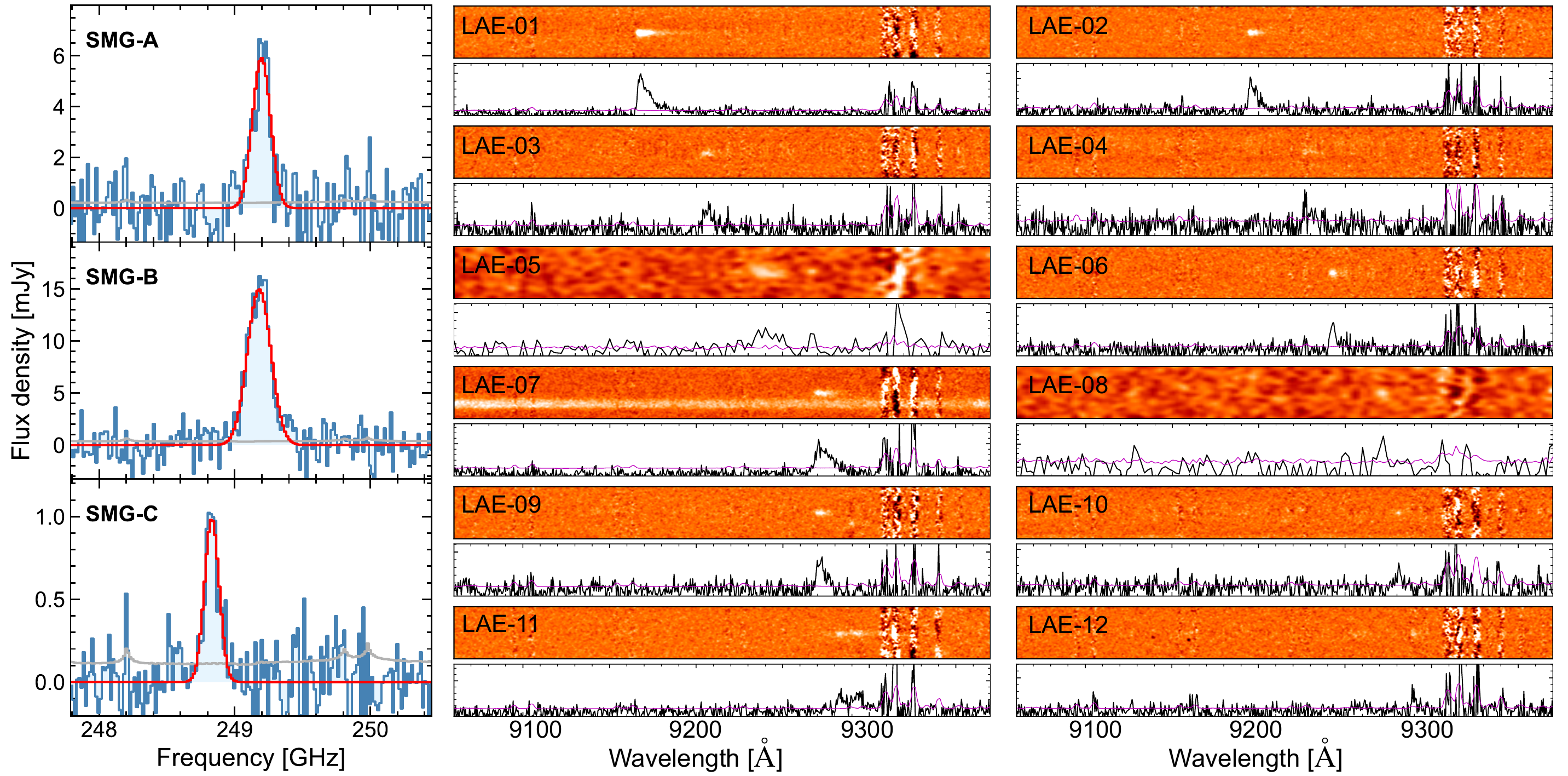}
\caption{ 
{\bf Left:} The [\ion{C}{2}] spectra (after continuum subtraction in the {\it uv} plane) of [\ion{C}{2}] emitters. 
Their redshifts are $z=6.6266\pm0.0001$, $6.6272\pm0.0001$, and $6.6379\pm0.0001$ for SMG-A, SMG-B, and SMG-C, respectively.
{\bf Middle and right:} The optical spectra of spectroscopically confirmed LAEs. The identifications of these LAEs are labeled as Arabic numerals in Figure \ref{fig:map}.
These galaxies have redshifts of $z=6.54-6.64$. Detailed information of the galaxies are listed in Table \ref{tbl:galaxy}. The spectra for LAE-05 and LAE-08 are from IMACS observations and the the remaining spectra are from DEIMOS observations.
}
\label{fig:spec}
\end{figure*}

\subsection{Optical spectroscopy} \label{subsec:spec}
To confirm the imaging selected LAE candidates and understand the contamination rate of our selection, we performed optical spectroscopic observations for a subset of the LAEs with Magellan/IMACS and Keck/DEIMOS. 
The Magellan/IMACS observations were carried out on March 2021. We used the IMACS f/2 camera with 300-line red-blazed grism and designed a dedicated multi-slit mask (slit mask ID is 4129 in the Magellan/IMACS mask database) for the observation. 
With a single mask, we observed 10 LAE candidates, some dropout-selected galaxies,  bright alignment stars and flux-calibration stars. All slits on the mask have a $1''$ slit width. 
The sky was clear with a seeing of $\sim0.8''-1.0''$ on the first night (2021 March 5) and $\sim0.6''-0.7''$ on the second night (2021 March 10). We obtained 11$\times$1200s exposures on the first night and 6$\times$1200s exposures on the second night. 
Following the standard procedures, we reduced IMACS data with the COSMOS2 pipeline \citep{COSMOS}. 
From these observations, we spectroscopically confirmed six LAEs at $z\sim6.6$ and one low-redshift [\ion{O}{3}] emitter at $z\sim0.85$. One object falls into a bad slit and the other two galaxies are undetected by the IMACS observations.

To obtain high quality spectra for IMACS confirmed LAEs and confirm more LAEs in the field, we also observed this field with Keck/DEIMOS. We designed three multi-slit masks with $1''$ slit width and used the 830G grating for the DEIMOS observations. We observed one mask on 2022 February 25 and two masks on 2022 February 26. We placed 8 LAE candidates on mask1, 4 LAE candidates on mask2, and 6 LAE candidates on mask3. 
We also added some dropout-selected galaxies and alignment stars as filler targets on these masks. 
The sky was clear on both nights and the seeing varied from $0.6''$ to $1.2''$ with a short period of $\sim1.5''$ seeing on the second night. We spent 6-hours of time on-source for mask1 and 3-hours each on the other two masks. 

The DEIMOS data were reduced with the PypeIt pipeline \citep{Pypeit_ascl, Pypeit} following the standard steps, including overscan subtraction, flat fielding, wavelength calibration, optimal extraction, flux calibration and telluric correction. 
Seven of the eight LAEs on mask1 were confirmed to be $z\sim6.6$ LAEs, with three of them previously confirmed with IMACS. 
The other LAE candidate was undetected because of its faintness (NB926=25.8 magnitude).
Masks mask2 and mask3 were observed for only 3-hours and only three (one of them was previously confirmed with IMACS) of the ten LAE candidates were confirmed to be $z\sim6.6$ LAEs. One LAE candidate is too close to the slit edge. The rest of the LAE candidates on mask2 and mask3 are either undetected or show very faint emissions and cannot be either confirmed or ruled out as $z\sim6.6$ LAEs. Future deeper spectroscopic observations are needed to confirm the nature of these LAE candidates. 

In total, the IMACS and DEIMOS observations convincingly confirmed 12 bright LAEs at $z\sim6.6$ and only one low-$z$ interloper at $z=0.85$. 
We can only measure redshift for $\sim$60\% of the LAE candidates but of these 92\% are indeed at the intended redshift $z\sim6.6$ suggesting that the narrow-band selection has low interloper contamination.
These bright LAEs have Ly$\alpha$ luminosities of $L_{\rm Ly\alpha}\simeq (0.6-3.6) \times10^{43}~{\rm erg~s^{-1}}$. The detailed properties of these spectroscopically confirmed $z\sim6.6$ LAEs are listed in Table \ref{tbl:galaxy} and the spectra of these LAEs are shown in Figure \ref{fig:spec}.

\begin{deluxetable*}{cccccccccrcr}
\tablecaption{Spectroscopically confirmed high-redshift galaxies in the protocluster.}\label{tbl:galaxy}
\setlength{\tabcolsep}{2pt}
\tablehead{\colhead{ID} & \colhead{Name} & \colhead{Redshift}& \colhead{Mask ID}& \colhead{NB926} & \colhead{$z$\tablenotemark{a}} & \colhead{$i$} & \colhead{$L_{\rm Ly\alpha}$} & \colhead{$L_{\rm [CII]}$}}
\startdata
     &             &              &                 & mag      & mag &mag&  $\rm 10^{43} erg~s^{-1}$   &  $10^{8}~L_\odot$ &  \\
\hline
LAE-01 & J091051.899$-$041703.94 & 6.541 & mask1,4129 & 24.36$\pm$0.05 & 25.46$\pm$0.10 & $>$27.81 & 3.57$\pm$0.87 & -- \\
LAE-02 & J091039.434$-$042258.45 & 6.563 & mask1          & 25.29$\pm$0.08 &            $>$27.19 & $>$27.73 & 1.34$\pm$0.09 & --  \\
LAE-03 & J091043.771$-$040755.72 & 6.572 & mask2          & 25.28$\pm$0.07 & 26.91$\pm$0.30 & $>$27.81 & 1.21$\pm$0.13 & --\\
LAE-04 & J091042.675$-$042828.98 & 6.589 & mask3          & 25.06$\pm$0.09 &            $>$27.17 & $>$27.73 & 0.76$\pm$0.16 & --  \\
LAE-05 & J091135.005$-$041744.02 & 6.597 & 4129            & 23.94$\pm$0.05 & 25.70$\pm$0.20 & $>$27.80 & 2.71$\pm$0.36 & --  \\
LAE-06 & J091022.398$-$040522.91 & 6.602 & mask2,4129 & 24.54$\pm$0.07 & 26.24$\pm$0.27 & $>$27.64 & 1.20$\pm$0.08 & -- \\
LAE-07 & J091101.294$-$041522.55 & 6.625 & mask1,4129 & 24.12$\pm$0.04 & 26.60$\pm$0.26 & $>$27.79 & 2.25$\pm$0.12 & -- \\
LAE-08 & J091122.389$-$035809.51 & 6.626 & 4129            & 24.71$\pm$0.10 & 25.74$\pm$0.17 & $>$27.75 & 0.78$\pm$0.12 & --  \\
LAE-09 & J091042.883$-$041734.68 & 6.626 & mask1         & 26.00$\pm$0.15 &            $>$27.17 & $>$27.82 & 1.24$\pm$0.10 & -- \\
LAE-10 & J091046.177$-$041747.74 & 6.634 & mask1         & 26.08$\pm$0.16 &            $>$27.24 & $>$27.86 & 0.65$\pm$0.10 & --  \\
LAE-11 & J091054.343$-$041452.80 & 6.641 & mask1,4129 & 24.66$\pm$0.06 & 26.98$\pm$0.34& $>$27.83& 1.86$\pm$0.30 & --  \\
LAE-12\tablenotemark{b} & J091053.671$-$041322.62 & 6.641 & mask1,4129 & 24.97$\pm$0.10 &           $>$27.19 & $>$27.82& 0.42$\pm$0.11 & -- \\ 
SMG-A & J091054.544$-$041355.52 &6.6266&   --                 & -- & -- & -- & -- &12.82$\pm$0.43  \\ 
SMG-B & J091053.598$-$041409.11 &6.6272&   --                 & -- & -- & -- & -- &42.43$\pm$0.73  \\
SMG-C & J091054.552$-$041406.99 &6.6379&   --                & -- & -- & -- & -- &1.76$\pm$0.16 
\enddata
\tablenotetext{a}{We adopt 3-$\sigma$ magnitude limits for undetected sources.}
\tablenotetext{b}{The Ly$\alpha$ line of LAE-12 was only detected from the DEIMOS data, limited by the depth of the IMACS observation.}
\end{deluxetable*}

\subsection{ALMA observation and [\ion{C}{2}] emitters} \label{subsec:alma}
To identify obscured galaxies in the quasar field, we also observed the quasar J0910--0414 with Atacama Large Millimeter/submillimeter Array (ALMA) C-3 in band 6 (Project: 2018.1.01188.S, PI: F. Wang). The on-source exposure time was 19 minutes. We identified two [\ion{C}{2}] emitting galaxies (SMG-A and SMG-B) from the single-pointing ALMA observation, indicating that we are seeing a significant small-scale overdensity of galaxies around this quasar. 
We then followed up this quasar field with ALMA/C-6 (Project: 2019.1.01216.S, PI: J. Yang) to reach a higher resolution and better sensitivity. The high resolution observations were obtained from 2021 May to 2021 June with a total on-source time of 383 minutes and revealed a third [\ion{C}{2}] emitting galaxy (SMG-C) interacting with the quasar itself. 

The ALMA data were reduced with the standard {\tt CASA} package \citep{CASA}.
In this letter, we focus on the identification of three [\ion{C}{2}] emitting galaxies, while the detailed ALMA data reduction and the characterization of the quasar host galaxy will be presented in a separate work (Yang J. et al., in prep).  
The two bright [\ion{C}{2}] emitting galaxies (SMG-A and SMG-B) are clearly seen in the moment zero map in panel c of Figure \ref{fig:map} and the third galaxy is merging with the quasar itself and can be seen in the the zoom-in plot in Figure \ref{fig:map}. The [\ion{C}{2}] spectra of galaxy SMG-A and SMG-B were extracted from the deep C-6 data cube with a  $1.0''$ diameter aperture centered at (09h10m54.544s, --04d13m55.515s) and (09h10m53.598s, --04d14m09.107s), respectively. 
To avoid contamination from the flux of the quasar, 
we extracted the spectra for the SMG-C with a smaller aperture ($0.3''$ diameter aperture ) centered at (09h10m54.552s, --04d14m06.994s). The [\ion{C}{2}] spectra of all these galaxies are also shown in Figure \ref{fig:spec} and the apertures used for spectral extraction are highlighted with yellow circles in panel {\bf b} and {\bf c} of Figure \ref{fig:map}.
The [\ion{C}{2}] luminosities of SMG-A, SMG-B, and SMG-C are in the range of $(1.8-42.5)\times10^8~L_\odot$.
It is interesting to note that all three [\ion{C}{2}] emitters are not detected in Subaru narrow-band imaging and $z$-band imaging, which indicates that these [\ion{C}{2}] emitters are likely dustier than the narrow-band selected LAEs.

\section{Results} \label{sec:cluster}
\subsection{Overdensity of galaxies}\label{subsec:density}
To measure the significance of the overdensity of LAEs in the protocluster, 
we measure the density of LAEs in two regions. First, we measure the LAE density in the outskirts of the Subaru HSC field. As shown in Figure \ref{fig:map}, the LAEs are clustered within the central $\sim14\times30$ arcmin$^2$ or $\sim35\times74$ cMpc$^2$ region. We choose to only use the region with $r>50$ cMpc \citep[i.e., radius bigger than the largest protocluster known in the EoR;][]{Jiang18} from the central quasar as a comparison field to measure the `background' LAE density. Second, we measure the LAE density within the central $\sim14\times30$ arcmin$^2$ region where the LAEs are clustered. 

\begin{figure*}
\centering
\includegraphics[width=0.49\linewidth]{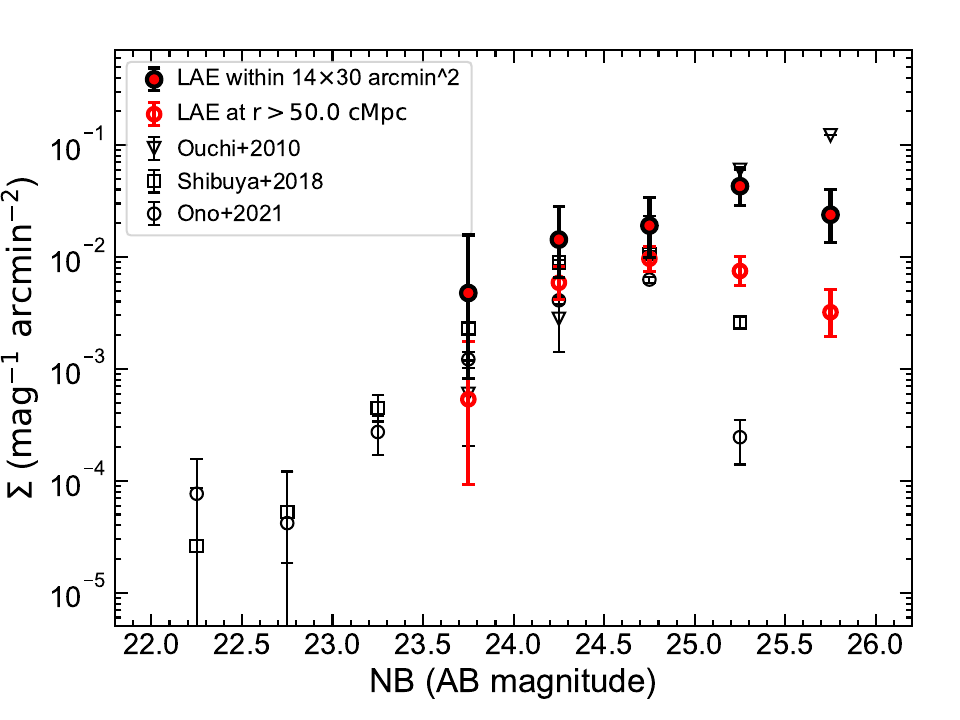}
\includegraphics[width=0.49\linewidth]{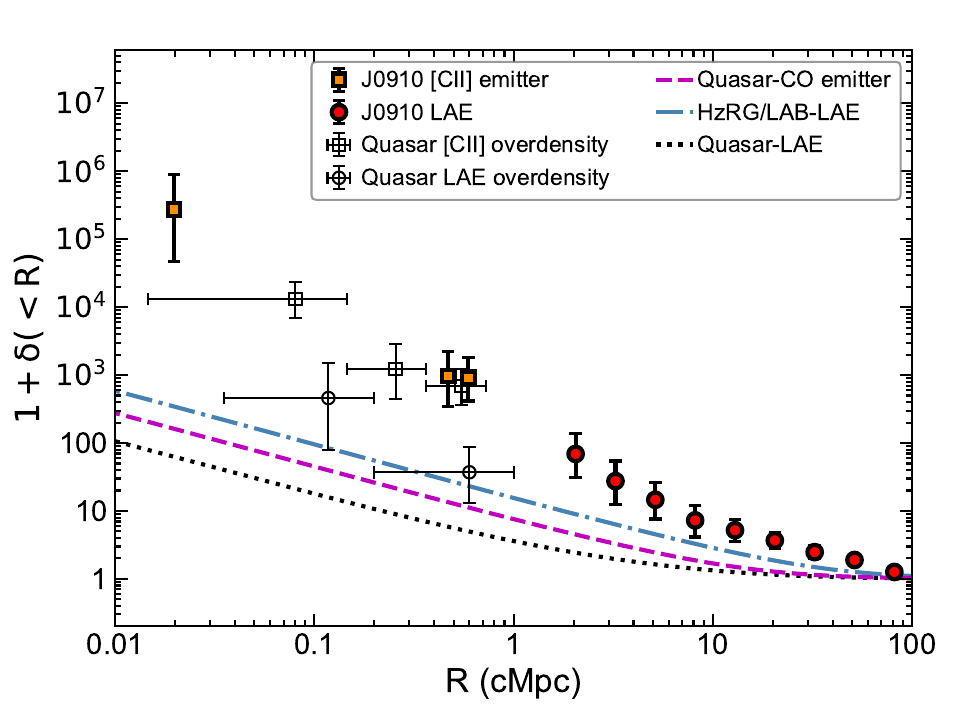}
\caption{\small 
\textbf{Left,}
 Surface density of LAE candidates as a function of narrow band magnitude.
 The density of LAE in the central $\sim14\times30$ arcmin$^2$ region (solid red points) is $4.3^{+1.1}_{-0.9}$ times higher than that measured in the outskirt ($r>50 ~{\rm cMpc}$) of our HSC field and that measured in other LAE surveys \citep{Ouchi10, Shibuya18, Ono21}. 
\textbf{Right,}
Cumulative overdensity profile of galaxies in quasar fields. The orange squares and red dots denote the overdensity profiles measured in J0910--0414 field for [\ion{C}{2}] emitters and LAEs, respectively. The black open squares and open circles denote the overdensity profiles of [\ion{C}{2}] emitters and LAEs measured from three quasar fields with [\ion{C}{2}] companion galaxies.
The black dotted line represents the expected overdensity profile of LAEs in quasar fields \citep{Shen07,Ouchi18,Garcia19}. The magenta dashed line denotes the exptected overdensity profile of CO emitters in quasar fields \citep{Garcia22}. The blue dot-dashed line shows the expected overdensity profile of LAEs in average protocluster fields \citep{Hennawi15}. 
}
\label{fig:density}
\end{figure*}

In Figure \ref{fig:density} we plot the surface density of LAEs as a function of their narrow band brightness. 
As shown in Figure \ref{fig:density}, the `background' LAE densities (red open circles) in our HSC field are very similar to that measured from other $z\sim6.6$ LAE surveys \citep{Ouchi10, Shibuya18, Ono21}, except at the very faint end close to  the survey depth limit. 
The surface density of LAEs in the central $\sim14\times30$ arcmin$^2$ region, however, is clearly higher than that estimated from both the outskirts of the HSC field and other studies in all magnitude bins. On average, the density in the central region (24 LAEs within 420 arcmin$^2$) is $4.3^{+1.1}_{-0.9}$ ($\delta=\frac{n}{\bar{n}}-1=3.3^{+1.1}_{-0.9}$) times higher than that derived from the outskirts (50 LAEs within 3740 arcmin$^2$) of the field. 
In particular, the overdensity of LAEs within the central $\lesssim2$cMpc is $\delta>30$, making J0910--0414 among the most extreme (in terms of both extent and overdensity) overdensities of LAEs known in the early Universe \citep[e.g.,][]{Jiang18}.  

To characterize the overdensity and the extent of the protocluster in detail, we also measure the cumulative overdensity profile of both LAEs and [\ion{C}{2}] emitters in this field. We measure the overdensity profile in a set of cylindrical volumes with logarithmically spaced radial bins and a velocity bin of $\Delta v=\pm1000~{\rm km~s^{-1}}$ \citep{Meyer22} and $\Delta v=\pm2250~{\rm km~s^{-1}}$ (the width of NB926 filter) for [\ion{C}{2}] emitters and LAEs, respectively. 
The background number of [\ion{C}{2}] emitters is estimated from the [\ion{C}{2}] cumulative number density with $L_{\rm [CII]}>10^{8.2}~L_\odot$ \citep[$5\sigma$ limit of our ALMA observation with an assumption of line width of FWHM=200 km s$^{-1}$;][]{Popping19, Uzgil21} while the background number of LAEs is directly measured from the outskirts ($r>50$ cMpc ) of our HSC observations. The overdensity profiles of [\ion{C}{2}] emitters and LAEs are shown as orange squares and red dots in Figure \ref{fig:density}, respectively. 
In this figure, we also show the overdensity profiles of three quasar fields with known [\ion{C}{2}] overdensities measured using both [\ion{C}{2}] emitters and LAEs by \cite{Meyer22} \footnote{The [\ion{C}{2}] density estimated from the [\ion{C}{2}] luminosity function \citep{Popping19} is underestimated by a factor of two in Figure 10 of \cite{Meyer22}. We corrected this factor in Figure \ref{fig:density}. }. 

To quantitatively characterize the overdensity profile in this protocluter, we further compare it with the overdensity profiles expected from protoclusters and quasar-galaxy cross-correlation functions in Figure \ref{fig:density}.
\cite{Shen07} measured a correlation length of $r_0^{\rm QQ}=24.3~h^{-1}~{\rm Mpc}$ for a fixed slope of $\gamma=2.0$ using a sample of $\sim1500$ quasars at $z>3.5$, which indicates that high redshift quasars are among the most highly clustered population in the Universe. 
\cite{Garcia19} re-fit the measurements in \cite{Shen07} by fixing $\gamma=1.8$ and derive $r_0^{\rm QQ}=22.3~h^{-1}~{\rm Mpc}$. 
\cite{Ouchi18} measured a correlation length of $r_0^{\rm GG}=2.66~h^{-1}~{\rm Mpc}$ (for a fixed slope of $\gamma=1.8$) for $z\sim6.6$ LAEs. This yields an expected quasar-LAE cross-correlation length of $r_0^{\rm QG}=\sqrt{r_0^{\rm QQ}, r_0^{\rm GG}}=7.7~h^{-1}~{\rm Mpc}$ ($\gamma=1.8$). The cumulative overdensity profile measured from the $r_0^{\rm QG}$ and $\gamma$ for quasar-LAE is plotted as a black dotted line in Figure \ref{fig:density}. Since there is no existing measurement of the auto-correlation length for [\ion{C}{2}] emitters, we show the overdensity profile (magenta dashed line) measured from quasar-CO emitters with $r_0^{\rm QG}=8.37~h^{-1}~{\rm Mpc}$ and $\gamma=1.8$ \citep{Garcia22} as a comparison. Furthermore, we also show the LAE overdensity profile (blue dot-dashed line) in protocluster fields at intermediate redshifts that is measured from a sample of high-$z$ radio galaxies (HzRg) and Ly$\alpha$ blobs (LAB) by \citet{Hennawi15} --- we used $\gamma=1.8$ and $r_0^{\rm QG}=20~h^{-1}~{\rm Mpc}$ to be consistent with $\gamma$ of other profiles shown in Figure \ref{fig:density}. 
It is clear that the overdensity in the J0910--0414 field is much higher than that expected from the quasar--galaxy cross-correlation function and the overdensity even exceeds the average protocluster by a factor of seven for $R\lesssim2~{\rm cMpc}$,  decreasing to an excess of $\sim3$ on scales of $R\simeq5$ cMpc, and exhibits a steeper profile. This indicates that the protocluster traced by quasar J0910--0414 is among the most overdense structures known in the early Universe.

\subsection{The mass of the protocluster}\label{subsec:mass}
To estimate the total present-day mass of this protocluster, we follow the widely used formula by assuming that this structure will collapse into a cluster at $z=0$ \citep[e.g.,][]{Chiang13}:
$M_{z=0}=(1+\delta_m)\bar{\rho}V$, where $\bar{\rho}=3.88\times10^{10}M_\odot~{\rm cMpc^{-3}}$ is the mean matter density of the Universe, $V$ is the volume of the protocluster, and $\delta_m$ is the mass overdensity. We estimate the volume of the protocluster using the yellow box ($\sim14\times30$ arcmin$^2$ or $\sim35\times74$ cMpc$^2$) shown in Figure \ref{fig:map} and a line-of-sight depth of  $36.9~ {\rm cMpc}$ which is derived from the redshift difference of the highest-$z$ LAE and lowest-$z$ LAE in the spectroscopic sample. We note that the volume is conservative because the boundary does not cover the whole structure and the line-of-sight depth is a lower limit since the sensitivity of NB926 filter drops dramatically for detecting LAEs at $z\gtrsim6.65$. 
The observed galaxy overdensity within this region is $3.3^{+1.1}_{-0.9}$ and the bias parameter for $z\sim6.6$ LAEs is measured to be $b=4.54\pm0.63$ \citep{Ouchi18}. 
We thus can determine $\delta_m$ through $1+b\delta_m = C(1+\delta)$, where $C$ is the correction factor for the redshift space distortion. 
Extrapolating the \cite{Chiang13} formula to $z\sim6.6$ and higher masses, we derive $C=0.84$, $\delta_m=0.57$ and the present-day mass $M_{z=0}=6.9^{+1.2}_{-1.4}\times10^{15}~M_\odot$ for $\delta=3.3^{+1.1}_{-0.9}$ at $z=6.63$, . This would make J0910--0414 the most massive protocluster known at $z>6$ \citep[e.g.,][]{Jiang18,Hu21} and at $z=0$ it would be three times more massive than the nearby Coma cluster \citep{Merritt87}. 

\subsection{Comparison with other galaxy overdensities traced by $z\sim6$ quasars}\label{subsec:protoclusters}
J0100+2802, the most luminous quasar known at $z>6$, has been found to be associated with a galaxy overdensity with $\sim20$ [\ion{O}{3}] emitters in its vicinity \citep{Kashino22}. However, these [\ion{O}{3}] emitters are much fainter than the galaxies identified in J0910--0414 field and we could not compare the two structures directly. 
Similarly, J0305--3150 resides in a structure with overdensities of both [\ion{O}{3}] emitters \citep{Wang23a} and LBGs \citep{Champagne23}. Nevertheless, the narrow-band imaging observation of J0305--3150 indicates that no bright candidate LAEs presents in the vicinity of J0305--3150, and the surface density of LAEs on large scales appears to be underdense compared to control fields \citep{Ota18}, making it a different kind of structure from J0910--0414. 
J0836+0054, a radio-loud quasar at a slightly lower redshift ($z\sim5.8$), resides in a rich structure with overdensity of bright LAEs \citep{Overzier22}. In particular, the three $z\sim5.8$ LAEs confirmed in J0836+0054 field have similar Ly$\alpha$ luminosities \citep{Bosman20} to the LAEs found in J0910--0414 field. In addition, the LAE overdensity in the central $R\lesssim3$ cMpc regions of both fields are comparable \citep{Overzier22}. This indicates that the two structures around J0836+0054 and J0910--0414 could be similar despite the lack of dusty galaxy observations in J0836+0054 field. 
J1030+0524, a well studied quasar at $z=6.3$, has been found residing in a structure with overdensities of both LBGs and LAEs at various scales \citep{Mignoli20}. Interestingly, most spectroscopically confirmed galaxies in J1030+0524 field are much fainter than the LAEs found in this work, indicating that the galaxy populations in these two structures could be different. 
Finally, J1526--2050, a luminous quasar at $z\sim6.6$, has been studied with both ALMA and MUSE observations. These observations indicates that J1526--2050 traces an Mpc-scale structure with both [\ion{C}{2}] emitters and LAEs in the quasar vicinity, making it a structure potentially similar to the structure around J0910--0414. However, the wide-field imaging and spectroscopic observations in the J1526--2050 field are lacking and the extend of this structure is still unclear. 
In summary, the structures traced by distant quasars exhibit significant diversity, and the current non-uniform observations make it challenging to directly compare their physical properties and evolutionary phases. Therefore, conducting systematic investigations of the Mpc-scale environment of distant quasars in a more consistent manner would be valuable for understanding the connection between quasars and their Mpc-scale surroundings, as well as the evolutionary phases of these large-scale structures.

\subsection{A double peaked LAE in the quasar vicinity} \label{sec:gal}
\begin{figure}
\centering
\includegraphics[width=1.0\linewidth]{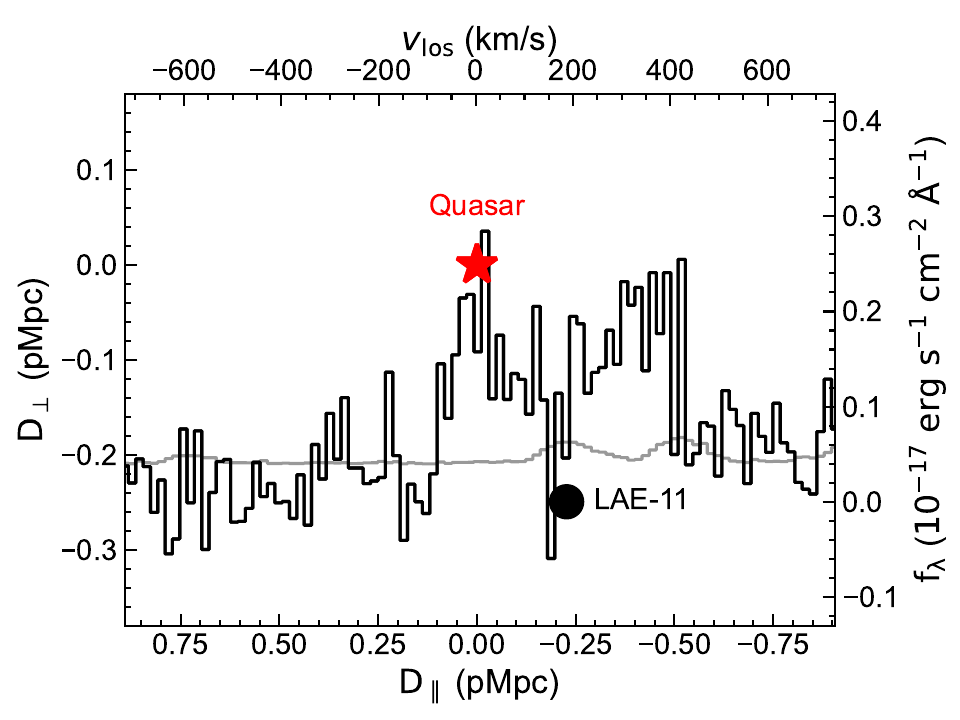}
\caption{ 
\textbf{A double peaked LAE identified in the highly ionized quasar vicinity.}
The red asterisk indicates the position of the quasar while the black circle highlight the relative position of the LAE-11 to the quasar. 
The black line shows the spectrum of the LAE and the gray line denotes the error vector. 
LAE-11 shows a widely separated double-peaked $\rm Ly\alpha$ line. The detection of the blue peak of the double peaked Ly$\alpha$ line in this galaxy indicates that the strong quasar radiation has ionized its vicinity. 
}
\label{fig:escape}
\end{figure}

Although the proximity zone size of the central quasar cannot be determined from the optical spectra because of the appearance of strong \ion{N}{5} broad absorption line on top of the quasar Ly$\alpha$ line \citep{Yang21}. Nevertheless, the typical proximity zone size of quasars with similar luminosity and redshift is $\sim$5 pMpc \citep{Eilers20a}. 
In Figure \ref{fig:escape}, we show the zoom-in spectra of the LAE-11, the nearest companion LAE at $\sim0.3$ pMpc from the quasar, and the relative position of the LAE-11 to the quasar. The line-of-sight proper distance and the perpendicular separation along the plane of sky between the quasar and the LAE-11 are $D_\parallel=0.18$ pMpc, and $D_\perp=0.25$ pMpc, respectively.
Interestingly, LAE-11, at a slightly higher redshift than the quasar, shows a widely separated double peak $\rm Ly\alpha$ line as shown in Figure \ref{fig:escape}. The peak separation of the $\rm Ly\alpha$ line is measured to be $320\pm22~{\rm km~s^{-1}}$ which implies a low ionizing photon escape fraction of $\sim7$\% by using the empirical fitting formula of \cite{Izotov18}. 
Since the blue peak of the Ly$\alpha$ line is only visible in a highly ionized environment, the presence of a double-peaked $\rm Ly\alpha$ line in LAE-11 provides a direct evidence that the quasar has ionized the gas in the intergalactic medium (IGM) around LAE-11 \citep[e.g.,][]{Bosman20}. This indicates that the quasar proximity zone size is $>0.3~{\rm pMpc}$. 
Additionally, we can constrain the quasar UV lifetime using galaxies in the proximity zones of quasars \citep[e.g.,][]{Bosman20} because the observed radiation field at a certain point in space is sensitive to the quasar radiation emitted at an earlier time with a time delay of 
\begin{equation}
    \Delta t = \frac{(D_\parallel^2+D_\perp^2)^{1/2}-D_\parallel}{c}
\end{equation}\label{equ:tq}
, where $c$ is the speed of light \citep{Bosman20}. 
Using Equation (4), we estimate the time delay to be $\Delta t=1.6\times10^6$ yr. 
Nevertheless, the redshift difference between the LAE-11 and the quasar is small ($\Delta z=0.005$ or $v=185~{\rm km~s^{-1}}$), we cannot distinguish the line-of-sight separation from the peculiar velocity of the galaxy around the quasar. In addition, the Ly$\alpha$ of high-redshift galaxies is usually redshifted \citep[e.g.,][]{Stark17}, which could make the line-of-sight velocity separation between the quasar and LAE-11 more uncertain. 
Taking these into account, we estimate the lower limit of quasar UV lifetime using $t_q > \frac{D_\perp}{c}=8.1\times10^5$ yr.
Future deeper spectroscopic observations of more galaxies around the quasar or at slightly higher redshifts than the quasar will allow us to study the ionizing property and the structure of quasar proximity zone \citep[e.g.,][]{Schmidt19}.    

\section{Summary} \label{sec:conclusion}
In this letter, we report the discovery of a protocluster anchored by a luminous quasar J0910--0414 at $z=6.63$ from wide-field Subaru/HSC imaging, Keck/DEIMOS and Magellan/IMACS optical spectroscopic, and deep ALMA observations.
The identification of this protocluster is secured by the discovery of a significant overdensity of both [\ion{C}{2}] emitters and LAEs in the field. 
The existence of three [\ion{C}{2}] emitters within just one ALMA pointing indicates that the density of [\ion{C}{2}] emitters in J0910--0414 field is three orders of magnitude times higher than that expected from blank field. The density of LAEs within the central $\sim14\times30$ arcmin$^2$ (or $\sim35\times74$ cMpc$^2$) region is $4.3^{+1.1}_{-0.9}$ times higher than that found in the outskirts of Subaru/HSC imaging, and the overdensity of LAEs increases to $\delta>30$ within the central two Megaparsecs. 
We measured the cumulative overdensity profile of galaxies in J0910--0414 field, which is much higher than that expected from the quasar--galaxy cross-correlation function from kpc-scale up to $\sim50$ Mpc-scale. The overdensity of LAEs even exceeds the average overdensity of protoclusters at $z\sim2$ by a factor of $\sim7$ for $R\lesssim2~{\rm cMpc}$ decreasing to an excess of $\sim3$ on scales of $R\simeq5$ cMpc. In addition, we estimated the present-day mass of this protocluster to be $M_{z=0}=6.9^{+1.2}_{-1.4}\times10^{15}~M_\odot$, three times more massive than the nearby Coma cluster, making J0910--0414 the most massive protocluster known at $z\gtrsim6$.

Our deep optical spectroscopy indicated a high purity of narrow-band selected LAE candidates and successfully confirmed 12 LAEs at $6.54<z\le6.64$ with only one low-$z$ interloper at $z=0.85$. Remarkably, we discover a double-peaked LAE in the vicinity of the central quasar. The presence of the blue peak Ly$\alpha$ line in this galaxy implies that the quasar vicinity is highly ionized by the quasar radiation and the quasar UV lifetime is larger than 0.8 Myr.

This work demonstrates that it is efficient to identify galaxy overdensities by targeting quasar fields with both ALMA and narrow-band observations. Future ALMA mosaic observations, wide-field narrow band imaging, and JWST observations of quasars in the EoR will allow us to not only identify galaxy overdensities in the EoR but also study the environmental dependence of galaxy evolution.  

\begin{acknowledgments}
We thank Yoshiaki Ono for providing their collected data for Ly$\alpha$ emitter surface densities. 
We thank Romain Meyer for providing the overdensity profiles of [\ion{C}{2}] and LAEs in three $z\sim6$ quasar fields and for insightful discussions about their measurements. 
We thank the referee for carefully reading the manuscript and providing great comments.
FW acknowledges support from NSF Grant AST-2308258.

This work was supported by a NASA Keck PI Data Award, administered by the NASA Exoplanet Science Institute. Data presented herein were obtained at the W. M. Keck Observatory from telescope time allocated to the National Aeronautics and Space Administration through the agency's scientific partnership with the California Institute of Technology and the University of California. The Observatory was made possible by the generous financial support of the W. M. Keck Foundation. 
This paper includes data gathered with the 6.5 meter Magellan Telescopes located at Las Campanas Observatory, Chile.
This research is based in part on data collected at the Subaru Telescope, which is operated by the National Astronomical Observatory of Japan. We are honored and grateful for the opportunity of observing the Universe from Maunakea, which has the cultural, historical, and natural significance in Hawaii.
The NB926 filter was supported by KAKENHI (26707006) Grant-in-Aid for Scientific Research (A) through the Japan Society for the Promotion of Science (JSPS).

This paper makes use of the following ALMA data: ADS/JAO.ALMA\#2018.1.01188.S and ADS/JAO.ALMA\#2019.1.01216.S. ALMA is a partnership of ESO (representing its member states), NSF (USA) and NINS (Japan), together with NRC (Canada), MOST and ASIAA (Taiwan), and KASI (Republic of Korea), in cooperation with the Republic of Chile. The Joint ALMA Observatory is operated by ESO, AUI/NRAO and NAOJ. The National Radio Astronomy Observatory is a facility of the National Science Foundation operated under cooperative agreement by Associated Universities, Inc.

The authors wish to recognize and acknowledge the very significant cultural role and reverence that the summit of Maunakea has always had within the indigenous Hawaiian community. We are most fortunate to have the opportunity to conduct observations from this mountain. 

\end{acknowledgments}

%

\vspace{5mm}
\facilities{ALMA, Keck:II (DEIMOS), Magellan:Baade (IMACS), Subaru (HSC)}


\software{astropy \citep{Astropy},  
Cosmos2 \citep{IMACS, COSMOS},
hscPipe \citep{Bosch18},
Matplotlib \citep{Matplotlib},
Numpy \citep{Numpy},
PypeIt \citep{Pypeit_ascl, Pypeit},
Scipy \citep{Scipy},
          }







\end{document}